\documentclass[12pt]{article}
\usepackage{epsfig}
\usepackage{url}
\makeatletter
\def\alt{\mathrel{\mathpalette\gl@align<}}
\def\agt{\mathrel{\mathpalette\gl@align>}}
\def\gl@align#1#2{\lower.6ex\vbox{\baselineskip\z@skip\lineskip\z@
\ialign{$\m@th#1\hfil##\hfil$\crcr#2\crcr\sim\crcr}}}
\makeatother

\begin{document}
\begin{flushright}
MI-TH-184\\
October, 2018
\end{flushright}
\vspace*{1.0cm}
\begin{center}
\baselineskip 20pt
{\Large\bf
Yukawa Unification with Four Higgs Doublets \\ in Supersymmetric  GUT} \vspace{1cm}

{\large
Bhaskar Dutta$^1$ and Yukihiro Mimura$^{2,3}$}
\vspace{.5cm}

$^1${\it
Department of Physics, Texas A\&M University,
College Station, TX 77843-4242, USA
}\\
%\vspace{.5cm}
$^2${\it
Department of Physics, National Taiwan University, Taipei 10617, Taiwan
}
\\
$^3${\it
Institute of Science and Engineering, Shimane University, Matsue 690-8504, Japan
}
\\

\vspace{.5cm}
%$^\dagger${\it
%}\\

\end{center}

\begin{center}{\bf Abstract}\end{center}

We  discuss the Yukawa coupling unification,  which can emerge in the grand unified  theory, in the context  of scenarios with more than one pair of Higgs doublet  since the current LHC constraint has become a problem for the Yukawa unification scenarios with just one pair of Higgs doublet. More than one pair of Higgs doublets can easily arise in missing partner mechanism which solves the doublet-triplet splitting problem. In such a scenario, the Yukawa unification occurs at a medium  $\tan\beta$ value, e.g., $\sim$ 30, which corresponds to much smaller threshold corrections compared to usual large $\tan\beta$ scenario for  $t-b-\tau$ unification in the context of SO(10) and $b-\tau$ unification in the context of SU(5) models. Further, we  show that an additional Higgs doublet pair lowers the sensitivity of the radiative symmetry breaking of the electroweak vacuum.

\thispagestyle{empty}

\bigskip
\newpage

\addtocounter{page}{-1}

\section{Introduction}
\baselineskip 18pt

The Standard Model (SM) is well established to describe the physics below the weak scale,
and the SM particle content is complete after the discovery of the Higgs boson whose mass is 125 GeV. However, 27\% abundance of the universe, origin of the electroweak scale, neutrino masses etc. are not explained in the SM. The minimal supersymmetric standard model (MSSM), one of the most elegant extension of the SM,  with origin in a grand unified model, e.g., SO(10)~\cite{so10}, has answers for all  these puzzles.

However,  there is no evidence of supersymmetry (SUSY) at the LHC  so far and this has generated constraints on the colored SUSY masses, e.g., squarks, gluino masses need to be $\geq$ 2 TeV~\cite{Sirunyan:2017kqq,Aaboud:2017vwy}. Similarly, the lower bounds on  non-colored SUSY masses have been kept on increasing. This situation has impacts on scenarios with predictions for lighter SUSY masses.  One such example is a scenario which  possesses unification of third generation Yukawa couplings motivated by the  grand unified theories \cite{Banks:1987iu,Ananthanarayan:1991xp,Dimopoulos:1991yz}. This scenario  is running into difficulty with the current LHC constraints on the sparticle masses~\cite{Gogoladze:2012ii,Poh:2015wta,Yamaguchi:2016oqz}. Since the unification of third generation Yukawa coupling can emerge in the context of minimal SO(10) unification scenarios~\cite{unification}, one wonders whether there is a way to circumvent this problem.

In addition,   the little hierarchy is becoming more fine-tuned with the  non-observation of SUSY. Since the SUSY breaking scale ($Q_S$) associated with stop mass is moving up, it becomes closer to the symmetry breaking scale ($Q_0$) where the Higgs squared mass turns negative by renormalization group equation (RGE). $Q_S$ is a dimensionful parameter while the smallness of $Q_0$ compared to the Planck scale is due to dimensionless parameters. The closeness of these two unrelated scales  defines the fine tuning of the little hierarchy which is elevating with the non-observation of SUSY particles~\cite{Dutta:2007az,Dutta:2016sva,Giudice:2006sn,Graham:2009gy,Baer:2012up,Baer:2016lpj}. 

Both problems seem to be ingrained in the choice of number of Higgs doublets in the low energy theory\footnote{
The phenomenological implications of the multi-pair of Higgs doublets
in the low energy SUSY models are discussed in \cite{Sakamura:1999ky,Escudero:2005hk}.}.
In the context of SO(10) or SU(5) GUT models,  more than one pair of Higgs doublets may exist in the full theory. The light pair of doublet arises by choosing smaller mass for one of the higher dimensional Higgs representations in the missing partner doublet-triplet splitting mechanism. However, more than one pair can easily be made  light as well.  %In such a situation, one can easily have 2 pairs of doublet being light without destroying any predction arising from the model.
We consider such a scenario and show that both problems can be solved, i.e.,  (i) Yukawa unification and (ii) less fine tuning in little hierarchy can be achieved in the context of 4 Higgs doublet (4HD) SUSY models arising from SO(10) or SU(5).  We show that the Yukawa coupling unification can be  realized for lower $\tan\beta$, for which the threshold corrections are quite small.

This paper is organized as follows. In section 2, we discuss missing partner model to understand the existence of more than one pair of Higgs doublets. In section 3, we describe $t-b-\tau$ unification and  in section 4, we discuss the Higgs potential and fine-tuning of little hierarchy. Section 5
contains our conclusion.

\section{Missing partner mechanism}

The ${\bf 126}+\overline{\bf126}$ representations ($\Delta+\bar\Delta$) in SO(10)
have three colored Higgs triplet reps $({\bf3},{\bf1},-1/3)$ (and three anti-triplet reps) under SM,
and two pairs of Higgs doublets $({\bf1},{\bf2},1/2)+({\bf1},{\bf2},-1/2)$.
If we adopt $\bf 210$ representation $\Phi$ to break SO(10),
there are one triplet and one pair of doublets.
Assuming that there are four ${\bf10}$-dimensional Higgs representations $H^a$ ($a=1,2,3,4$),
we obtain the Higgs triplet and doublet fields as
\begin{eqnarray}
{\cal H}_T &=& (\Delta_T^{(6,1,1)},\bar \Delta_T^{(6,1,1)},\bar \Delta^{(10,1,3)}_T,\Phi_T,H_T^1,H_T^2,H_T^3,H_T^4), \\
{\cal H}_{\bar T} &=& (\bar\Delta_{\bar T}^{(6,1,1)},\Delta_{\bar T}^{(6,1,1)},\Delta_{\bar T}^{(10,1,3)},\Phi_{\bar T},
H_{\bar T}^1,H_{\bar T}^2,H_{\bar T}^3,H_{\bar T}^4),\\
{\cal H}_u &=& (\Delta_u,\bar \Delta_u,\Phi_u,H_u^1,H_u^2,H_u^3,H_u^4), \\
{\cal H}_d &=& (\Delta_d,\bar \Delta_d,\Phi_d,H_d^1,H_d^2,H_d^3,H_d^4),
\end{eqnarray}
and the mass terms
\begin{equation}
({\cal H}_T)_i (M_T)_{ij} ({\cal H}_{\bar T})_j + ({\cal H}_u)_i (M_D)_{ij} ({\cal H}_{d})_j.
\end{equation}
The mass matrices are written as
\begin{eqnarray}
M_T = \left(
\begin{array}{cc}
 A_T^{(4\times 4)} & B_T^{(4\times4)} \\
 C_T^{(4\times 4)} & D_T^{(4\times4)}
\end{array}
\right),
\qquad
M_D = \left(
\begin{array}{cc}
 A_D^{(3\times 3)} & B_D^{(4\times3)} \\
 C_D^{(3\times 4)} & D_D^{(4\times4)}
\end{array}
\right).
\label{AB}
\end{eqnarray}
The matrices $A_T$ and $A_D$ are determined by the masses of $\Delta$ and $\Phi$ and their GUT-scale vevs,
which depend on the SO(10) symmetry breaking vacua\footnote{The minimal missing partner SO(10) model used in the Ref.~\cite{Babu:2006nf} uses {\bf10}, {\bf120}, {\bf126}, $\overline{\bf126}$ and $\bf210$. We have more doublets and triplets arising from four {\bf10}s compared to one {\bf10} and one {\bf120} which causes  the dimensions for the matrices in Eq.(\ref{AB}) to be different compared to the minimal model. However, the final conclusion is independent of the choice of a particular model.}.
The matrices $B_{T,D}$ and $C_{T,D}$ depend on the Higgs coupling $H^a \Delta \Phi$ and the GUT scale vevs.  
The matrices $D_D$ and $D_T$ are obtained by the mass term of {\bf10}-dimensional Higgs fields.
If the mass of ${\bf10}$ are suppressed then
one linear combination of the Higgs doublets remain light (at weak scale)
while all the other linear combinations of doublets and triplets are massive at the GUT scale. The smallness of mass can be  due to supersymmetry breaking  which can arise out of a term $ {\bf10}\ {\bf10} X/M_{\rm pl}$ in the Kahler potential where $X$ is  a SM singlet. Now  a non-zero $ \langle F_X\rangle$ allows us to get ${\bf10} \ {\bf10}\langle F_X \rangle/{M_{\rm pl}} $ in the superpotential which leads to $M_s {\bf 10}\ {\bf 10}$ where $M_s \sim$ weak scale. This is similar to the origin of the weak scale $\mu$ term \cite{Giudice:1988yz}. 
The Yukawa interactions to generate the fermion masses
are given as
\begin{equation}
W_Y = h^a_{ij} \psi_i \psi_j H^a + f_{ij} \psi_i \psi_j \bar\Delta.
\end{equation}
The charged fermion Yukawa matrices are given by a linear combination of $h^a$
since the mixing of $\bar\Delta_{u,d}$ in the light Higgs doublets are tiny under the assumption above.
The left- and right-handed Majorana neutrino masses can be generated by the $f$ coupling.
By investigating $M_T^{-1}$, one finds that the $f$ coupling does not contribute to the proton decay amplitudes
and the dimension-five operators $C_{L,R}^{ijkl}$ are the linear combination of $h^a_{ij} h^b_{kl}$~\footnote{In the missing partner mechanism, either one  or no triplet is light. In the case when one triplet is light while the others are heavy, which  may emerge in the missing partner  mechanism, the ${\overline{\bf126}}$ triplet component vanishes in the lighter colored triplet field content~\cite{Dutta:2004zh}.}.
Therefore, compared to the minimal SO(10) model, though the predictivity of the neutrino masses and mixings is lost,
the proton decay suppression is easier to be realized (in type II seesaw) by choosing the hierarchy pattern in $h^a$
(e.g., $h^1$ is a nearly rank-1 matrix, which gives top, bottom and tau Yukawa couplings, and 1st and 2nd generation
masses and CKM mixings are generated by the other $h^a$).
Surely, in this naive choice, the Georgi-Jarskog relations are not obtained.
Instead of requiring four ${\bf 10}$ Higgs fields, by adopting one ${\bf120}$ representation (which contains
two triplets and two pairs of doublets) and two ${\bf 10}$ fields\footnote{
We note that the ${\bf 120}$ contribution can violate the quark-lepton mass unification
and the masses of charged-leptons and down-type quarks can be fit.
However, observed up- and down-type quark masses cannot be fit
if only one {\bf 10} and $\bf 120$ couple to fermions by renormalizable terms \cite{Lavoura:2006dv}.
In order to fit the charged fermion masses, one needs two ${\bf10}$'s or non-renormalizable couplings \cite{Babu:2006nf}.
}, one can realize the same situation
where only one pair of doublets is light\footnote{
More general description of the missing partner mechanism in SO(10) GUT
can be found in Ref.\cite{Babu:2011tw}.
} and the Georgi-Jarskog relations can be realized~\cite{Dutta:2004zh}.

\section{$t-b-\tau$ Unification}

In the context of a minimal SO(10) model, the doublet-triplet splitting arises just by cancellation in the determinant of the doublet mass matrix,
and one of the linear combination is fine-tuned to be light.
In the missing partner doublet realization of the doublet-triplet splitting, on the other hand, the lightness of one pair of doublets is realized by the smallness of the
mass of the ${\bf10}$ (and ${\bf 120}$) Higgs representations, and in principle, there is no strong reason that
only one pair of doublets is light since it just depends on the number of ${\bf 10}$-dimensional Higgs fields.
It is possible that multi-pair of Higgs doublets can be light in this scenario,
which is true in the missing partner mechanism also in SU(5) and flipped-SU(5).

Here, let us consider the possibility that two pairs of Higgs doublets (totally, four Higgs doublets) remain light.
There are two possibility depending on the number of excess of the triplet $({\bf3},{\bf1},-1/3)$ compared to $({\bf1},{\bf2},1/2)$:
\begin{enumerate}
\item
Two pairs of doublets are light, and one triplet (and one anti-triplet) Higgs is light.
\item
Two pairs of doublets are light, and no triplet Higgs is light.
\end{enumerate}
In the case 1, to avoid rapid proton decay, the Yukawa interaction to the fermions of the Higgs triplet needs to be very tiny
 (by the discrete symmetry or anomalous U(1) symmetry).
Then, the Yukawa coupling of one of the linear combination of the Higgs doublets is absent.
In the case 2, both two linear combination of the doublets can couple to the fermions
and there are new FCNC sources in the Yukawa interaction.
Surely, in the case 2, the gauge coupling unification in MSSM is destroyed explicitly (though it can be restored by the GUT-scale or intermediate-scale threshold corrections).

We consider the consequence of the case 1\footnote{As we have mentioned,
the proton decay suppression can be more easily realized 
if only one ${\bf10}$ Higgs coupling generates 3rd generation masses
and the others generates the masses of 1st and 2nd generations and the generation mixings.
If one chooses so, the following discussion can be the same even in case 2 in principle.
In the case 2, the additional Higgs couplings to the 1st and 2nd generations
can induce new FCNC, which can be a source of lepton flavor non-universality,
in the similar way to the non-SUSY general (so called type III) two Higgs doublet model.
}.
Denoting that the linear combination of the Higgs doublets which couples to fermions as
$\hat H_{1u}$ and $\hat H_{1d}$ and the other combinations as $\hat H_{2u}$ and $\hat H_{2d}$ (we call this as Yukawa-basis),
we obtain the Yukawa terms (below the GUT scale) :
\begin{equation}
W = Y_u^{ij} q_i u^c_j \hat H_{1u} + Y_d^{ij} q_i d^c_j \hat H_{1d} + Y_e^{ij} \ell_i e^c_j \hat H_{1d}.
\end{equation}
The $\mu$-term and the SUSY breaking Higgs mass terms are given as
\begin{equation}
W = \hat\mu_{ij} \hat H_{iu} \hat H_{jd},
\end{equation}
and 
\begin{equation}
V_{\rm soft} = (\hat b_{ij} \hat H_{iu} \hat H_{jd} + c.c.) + \hat m^2_{H_u}{}_{ij} \hat H_{iu}^\dagger \hat H_{ju} + \hat m^2_{H_d}{}_{ij} \hat H_{id}^\dagger \hat H_{jd}.
\end{equation}
Via RGE (with a large $Y_u^{33}$), $m^2_{H_u}{}_{11}$ becomes negative and the electroweak symmetry is broken.
Not only $\hat H^0_{1u,d}$ but also $\hat H^0_{2u,d}$ acquires vevs
(denote them as $v_{iu}$ and $v_{id}$).
We define a new basis (called as vev-basis):
\begin{equation}
\left( 
 \begin{array}{c}
  H_{1u} \\
  H_{2u}
 \end{array}
\right)
= \left(
\begin{array}{cc}
 \cos\zeta_u & \sin\zeta_u \\
 -\sin\zeta_u & \cos\zeta_u
\end{array}
\right)
\left( 
 \begin{array}{c}
  \hat H_{1u} \\
  \hat H_{2u}
 \end{array}
\right),
\quad
\left( 
 \begin{array}{c}
  H_{1d} \\
  H_{2d}
 \end{array}
\right)
= \left(
\begin{array}{cc}
 \cos\zeta_d & \sin\zeta_d \\
 -\sin\zeta_d & \cos\zeta_d
\end{array}
\right)
\left( 
 \begin{array}{c}
  \hat H_{1d} \\
  \hat H_{2d}
 \end{array}
\right),
\end{equation}
where $\tan\zeta_u = v_{2u}/v_{1u}$
and $\tan\zeta_d = v_{2d}/v_{1d}$,
so that
$\langle H_{2u}^0 \rangle = \langle H_{2d}^0 \rangle = 0$,
$\langle H_{1u}^0 \rangle = v_u$,
and
$\langle H_{1d}^0 \rangle = v_d$.
We denote $v_u = \sqrt{v_{1u}^2+ v_{2u}^2}$
and $v_d = \sqrt{v_{1d}^2+ v_{2d}^2}$.
As usual, we define $\tan\beta = v_u/v_d$,
and $v=\sqrt{v_u^2+v_d^2}$ is fixed by the gauge boson masses.
The Yukawa terms (in the vev-basis) are
\begin{eqnarray}
W &=& Y_u^{ij}  q_i u^c_j (\cos\zeta_u H_{1u} - \sin\zeta_u H_{2u}) + Y_d^{ij} q_i d^c_j  (\cos\zeta_d H_{1d} - \sin\zeta_d H_{2d})   \\
&&+ Y_e^{ij} \ell_i e^c_j (\cos\zeta_d H_{1d} - \sin\zeta_d H_{2d}) , \nonumber
\end{eqnarray}
and the fermion mass matrices
are
\begin{equation}
M_u = Y_u v_u \cos\zeta_u,
\quad
M_d = Y_d v_d \cos\zeta_d,
\quad
M_e = Y_e v_d \cos\zeta_d.
\end{equation}
We find that the RGE running of the top, bottom and tau Yukawa couplings (whose description is easier in Yukawa-basis)
 for $\cos\zeta_u \simeq1 $, $\tan\zeta_d \sim1$ and $\tan\beta\sim 30$
resembles the running in MSSM for $\tan\beta \simeq 50$.
In other words, for $\tan\beta\alt 35$ in the MSSM, the bottom Yukawa coupling 
 is small and the RGE running  is governed by the loop diagram arising from the gauge interaction $y_b^2 \ll g_3^2$.
On the other hand,
since $y_b = Y_d^{33} \cos\zeta_d$, $Y_d^{33}$ can be $\sim1$, the Yukawa interaction can contribute to the RGE evolution of bottom mass even if $y_b$ is small.
This freedom can make the top, bottom and tau Yukawa unification possible for   $\tan\beta\sim 30$
if the weak scale threshold correction is small.

We plot the RGE running of the couplings $Y_{u,d,e}^{33}$ in Fig.1 for different values of $\cos\zeta_u$, assuming that the third generation Yukawa couplings are unified
at $M_U = 2\times 10^{16}$ GeV.
In Fig.2, we plot the bottom-tau mass ratio at $M_Z$ leaving out the weak scale threshold corrections
as a function of $\cos\zeta_u$.

\begin{figure}[htb]
\center
\includegraphics[width=8cm]{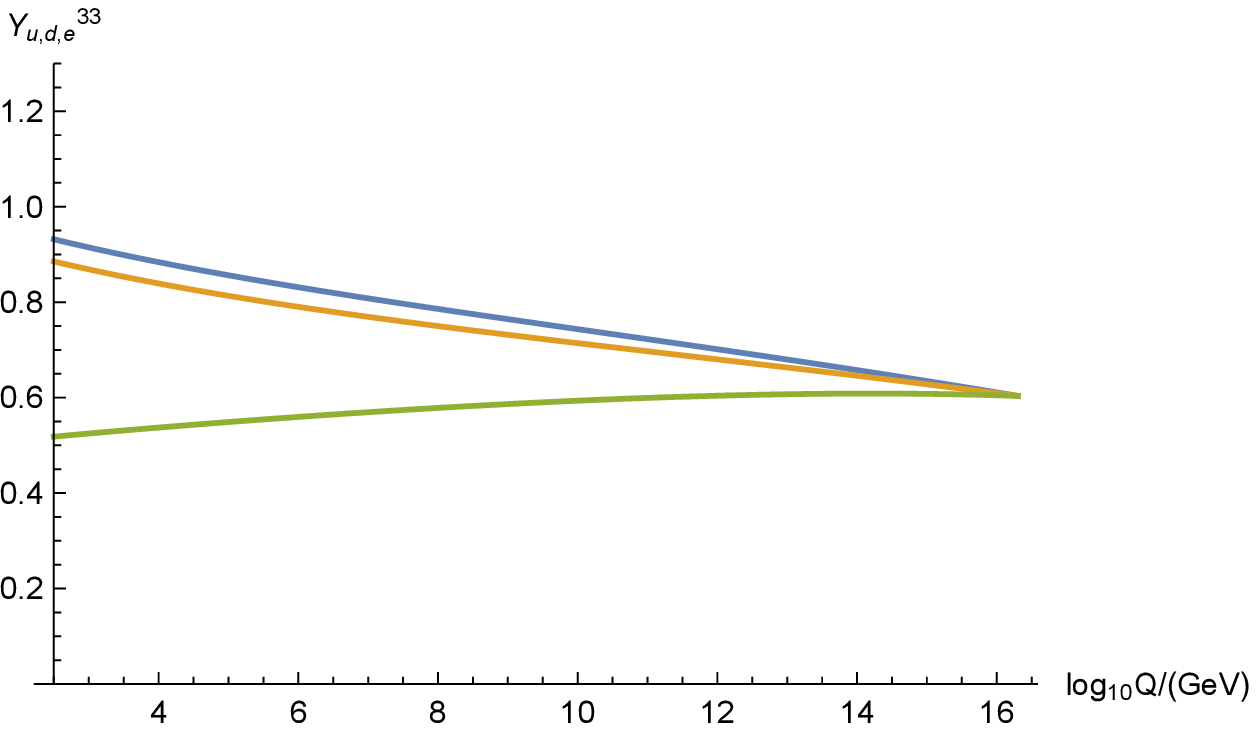}
\includegraphics[width=8cm]{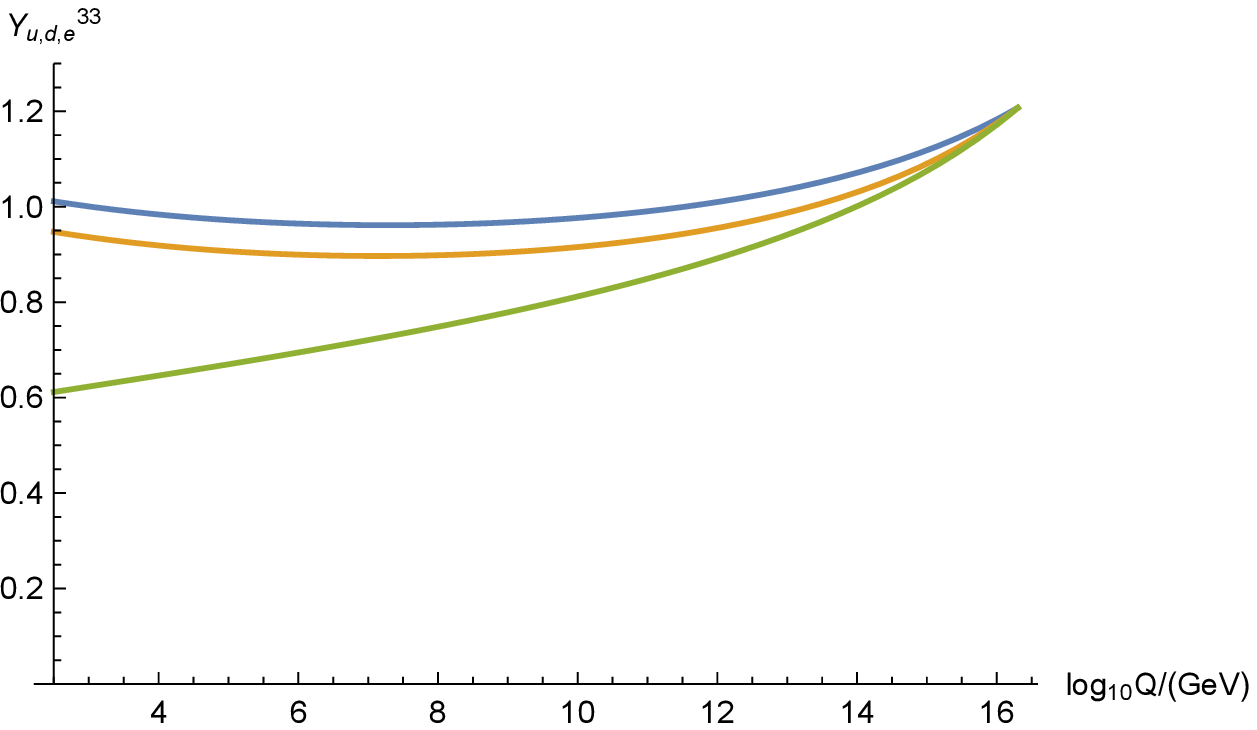}
\caption{The RGE running of the couplings for $\cos\zeta_u = 1$ (left), $\cos\zeta_u = 0.92$ (right).
From top to bottom, the couplings correspond to $Y_u^{33}$, $Y_d^{33}$ and $Y_e^{33}$.
}
\end{figure}

\begin{figure}[htb]
\center
\includegraphics[width=8cm]{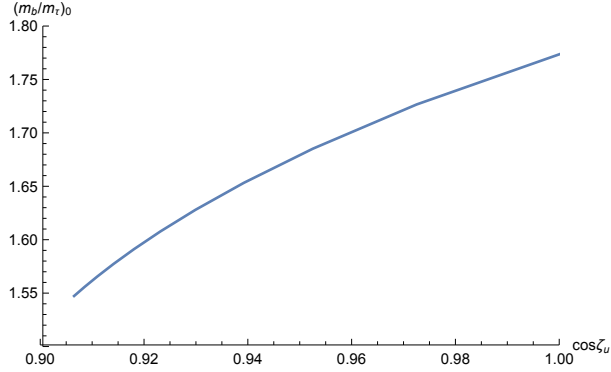}
\caption{The bottom-tau mass ratio $(m_b/m_\tau)_0$ at $M_Z$ without the weak scale threshold corrections, assuming the Yukawa unification. The RGE running of $Y_{d,e}^{33}$ depends on $Y_u^{33} = m_t/(v_u \cos\zeta_u)$,
and thus the ratio depends on $\cos\zeta_u$. For a reference value,
the running bottom-tau mass ratio at $Z$ boson mass scale is $m_b/m_\tau  \simeq 1.63$ \cite{Tanabashi:2018oca}.
}
\end{figure}

In the MSSM, it has been discussed that the bottom and tau unification is possible if $\tan\beta \sim2$ or $\tan\beta \sim 50$.
For $\tan\beta\sim2$, the top Yukawa coupling is large and it can contribute to the RGE running of bottom Yukawa coupling,
but, at present $\tan\beta\sim2$ is excluded due to the 125 GeV Higgs mass.
For $\tan\beta\sim50$, the finite corrections and the TeV scale threshold corrections are large and it is difficult to realize
the bottom-tau unification for the current bounds on the SUSY mass spectrum.
Actually, the finite correction of $q_3 b^c H_u^*$ is important for large $\tan\beta$:
\begin{equation}
m_b = y_b v_d \left(1 + X(\mu, A_t, m_{\tilde t}, m_{\tilde g}, m_{\tilde b})  \tan\beta\right).
\end{equation}
Naively, for a stop mass $\sim 2-3$ TeV, $A_t$ has to be large (for the 125 GeV Higgs) which makes  the chargino contribution is large 
%(${{\Delta m_b^{\tilde{\chi}^\pm_i}}\over{m_b}}\simeq {\lambda_t^2\over{16 Pi^2}} \mu (A_t\, tan\beta-\mu)I(mu^2, m^2_{\tilde{t}_1}, m^2_{\tilde{t}_2}$),
($X^{\tilde \chi^{\pm}} \simeq \frac{\lambda_t^2}{16\pi^2} \frac{\mu (A_t -\mu \cot\beta)}{m_{\tilde t_1}^2-m_{\tilde t_2}^2} I(\mu^2, m_{\tilde t_1}^2, m_{\tilde t_2}^2)$),
and the gluino contribution is large if $m_{\tilde g}$ is large  
%(${{\Delta m_b^{\tilde{g}_3}}\over{m_b}}\simeq {8\over{3}}{g_3^2\over{16 Pi^2}} m_{\tilde{g}} (\mu\, tan\beta-A_b)I(\tilde{g}^2, m^2_{\tilde{b}_1}, m^2_{\tilde{b}_2}$).
($X^{\tilde g} \simeq \frac{g_3^2}{12\pi^2} \frac{m_{\tilde g} (A_b\cot\beta -\mu)}{m_{\tilde b_1}^2-m_{\tilde b_2}^2} I(m_{\tilde g}^2, m_{\tilde b_1}^2, m_{\tilde b_2}^2)$).

In 4HD case, there are additional contributions to $X$ compared to MSSM if $\mu_{12}$ (in vev-basis) is not zero.
In the chargino loop (Higgsino component), $\mu_{12}$ can directly contribute in the Higgsino propagator.
In the gluino loop, there is a term $y_b (\mu_{11} - \mu_{12} \tan\zeta_d) \tilde q_3 \tilde b^c H_{1u}^*$
in the $F$-term, $|\partial W/\partial H_{id}|^2$.
Therefore, if $\mu_{12}$ is small, the contribution to $X$ is similar to MSSM and the finite correction is not sizable 
for $\tan\beta$ which is not so large.

%In summary, in MSSM (2HD), in the case that the RGE running is suitable for the bottom-tau unification for large $\tan\beta$,
In summary, in the context of MSSM with 2HD, RGE running is important  for the bottom-tau unification for large $\tan\beta$ but 
the large finite correction associated with the non-observation of SUSY masses destroys the realization of the bottom-tau unification.
In 4HD, however, the suitable RGE running can be realized even for  $\tan\beta$ which somewhere in the middle where the finite corrections are not sizable,
and as a result, top, bottom and tau Yukawa unification is possible in a simple manner.
In the missing partner mechanism for the doublet-triplet splitting, the existence of two pairs of Higgs doublets with masses around the weak scale is not at all unnatural. 
%does not appear to be   improbable.

\section{Minimization of the Higgs potential}

In 2HD case, 
the Higgs potential of the neutral Higgs vevs is
\begin{equation}
V = m_1^2 v_d^2 + m_2^2 v_u^2 + 2m_3^2 v_d v_u + \frac{g_Z^2}8 (v_d^2-v_u^2)^2,
\end{equation}
where $m_1^2 = m_{H_d}^2 + \mu^2$ and $m_2^2 = m_{H_u}^2 + \mu^2$.
The $Z$ boson mass (at tree-level) 
is written as
\begin{equation}
\frac{M_Z^2}{2} \cos^22\beta = - 
\left(
\begin{array}{cc}
 \sin\beta &\cos\beta
\end{array}
\right)
\left(
\begin{array}{cc}
 m_2^2 & m_3^2 \\
 m_3^2 & m_1^2
\end{array}
\right)
\left(
\begin{array}{c}
 \sin\beta \\ \cos\beta
\end{array}
\right).
\end{equation}
The other minimization condition gives 
\begin{equation}
m_3^2 = - \frac12 (m_1^2+m_2^2) \sin2\beta.
\end{equation}
The symmetry breaking condition (which is equivalent to $M_Z^2>0$) is $m_1^2 m_2^2 - m_3^4 < 0$.
For a large $\tan\beta$, we obtain 
$M_Z^2 \simeq -2 m_2^2$ and a cancellation between $\mu^2$ and $- m_{H_u}^2$ is needed (if $|m_{H_u}^2|$ is large
for a given boundary condition of SUSY breaking).
It is often said that a smaller $\mu$ is preferable for ``Natural SUSY" due to the tree-level relation.
However, the Higgs mass parameters run by RGEs, and it is still unnatural if 
the RGEs give a large logarithmic correction to the mass parameters near the minimization scale
(where the 1-loop correction of the scalar potential $\Delta V$ gives small derivatives
$\partial \Delta V/\partial v_u \approx \partial \Delta V/\partial v_d \approx 0$).
For example, if $m_{H_u}^2$ runs rapidly, the radiative electroweak symmetry breaking is still sensitive to the
SUSY breaking parameters 
(even if one tunes $|m_{H_u}^2|$ to be small at a scale). 
Such a situation can be expressed by equations as follows:
By Tailor expansion around the scale $Q_0$, the $Z$ boson mass relation can be 
expressed as
\begin{equation}
\frac{M_Z^2}{2} \cos^22\beta \simeq  
\left(
\begin{array}{cc}
 \sin\beta &\cos\beta
\end{array}
\right)
\left(
\begin{array}{cc}
 \frac{dm_2^2}{d\ln Q} & \frac{dm_3^2}{d\ln Q} \\
 \frac{dm_3^2}{d\ln Q} & \frac{dm_1^2}{d\ln Q}
\end{array}
\right)
\left(
\begin{array}{c}
 \sin\beta \\ \cos\beta
\end{array}
\right) \ln \frac{Q_0}{Q_S},
\end{equation}
where $Q_0$ is the symmetry breaking scale satisfying $m_1^2 m_2^2 - m_3^4=0$,
and $Q_S$ is the minimization scale, which is roughly same as the geometric average of the stop masses.
The RGEs of $m_2^2$, $m_1^2$ and $m_3^2$ are given as
\begin{eqnarray}
(4\pi)^2 \frac{dm_2^2}{d\ln Q} &=& 
6 (y_t^2 (m_{\tilde q_{3L}}^2 + m_{\tilde t_R}^2) + A_t^2) + 6y_t^2 m_2^2 - 6 g_2^2 M_2^2 - 2g'^2 M_1^2 \\
&&+ (6y_b^2 + 2y_\tau^2 - 6 g_2^2 - 2g'^2) \mu^2, \nonumber\\
(4\pi)^2 \frac{dm_1^2}{d\ln Q} &=& 
6 (y_b^2 (m_{\tilde q_{3L}}^2 + m_{\tilde b_R}^2) + A_b^2) 
+ 2 (y_\tau^2 (m_{\tilde \ell_{3L}}^2 + m_{\tilde \tau_R}^2) + A_\tau^2) 
\\
&& + (6y_b^2+2y_\tau^2) m_1^2 - 6 g_2^2 M_2^2 - 2g'^2 M_1^2 \nonumber\\
&&+ (6y_t^2 - 6 g_2^2 - 2g'^2) \mu^2, \nonumber\\
(4\pi)^2\frac{dm_3^2}{d\ln Q} &=& 
(3y_t^2 + 3y_b^2 +y_\tau^2-3g_2^2 - g'^2) m_3^2 \\
&&+ 6 g_2^2 M_2 \mu + 2 g'^2 M_1 \mu
+6 \mu A_t y_t +6\mu A_b y_b +2 \mu A_\tau y_\tau. \nonumber
\end{eqnarray}
One can find that 
 $\ln Q_0/Q_S$ to needed to be tuned to be small (irrespective of the smallness of $\mu$) 
if the stop masses and $A_t$ are large. 

Let us examine the ``naturalness" of the little hierarchy in the case of 4HD.
The Higgs potential in 4HD (in the Yukawa-basis) is given as
\begin{equation}
V = \left(
\begin{array}{cccc}
 v_{1u}  & v_{2u} & v_{1d} & v_{2d}
\end{array}
\right)
M_0^2\left(
\begin{array}{c}
v_{1u}  \\ v_{2u} \\ v_{1d} \\ v_{2d}
 \end{array}
\right) + \frac{g_Z^2}8 (v_{1u}^2 + v_{2u}^2- v_{1d}^2 - v_{2d}^2)^2,
\end{equation}
where
\begin{equation}
M_0^2
= \left(
 \begin{array}{cccc}
   m_{u11}^2 & m_{u12}^2 & \hat b_{11} & \hat b_{12} \\
   m_{u12}^2 & m_{u22}^2 & \hat b_{21} & \hat b_{22} \\
   \hat b_{11} & \hat b_{21} & m_{d11}^2 & m_{d12}^2 \\
   \hat b_{12} & \hat b_{22} & m_{d12}^2 & m_{d22}^2
 \end{array}
\right),
\end{equation}
and
\begin{eqnarray}
m_{u11}^2 &=& (\hat m_{H_u}^2)_{11} + \hat \mu_{11}^2 + \hat \mu_{12}^2, \\
m_{u12}^2 &=& (\hat m_{H_u}^2)_{12} + \hat \mu_{11} \hat \mu_{21} + \hat \mu_{12} \hat\mu_{22}, \\
m_{d11}^2 &=& (\hat m_{H_d}^2)_{11} + \hat \mu_{11}^2 + \hat \mu_{21}^2, 
\end{eqnarray}
and so on.
The minimization conditions of the tree-level potential can be written as
\begin{equation}
\frac{M_Z^2}{2} (-\cos2\beta) 
\left(
\begin{array}{c}
v_{1u}  \\ v_{2u} \\ -v_{1d} \\ -v_{2d}  \end{array}
\right)
= - M_0^2 
\left(
\begin{array}{c}
v_{1u}  \\ v_{2u} \\ v_{1d} \\ v_{2d}  \end{array}
\right).
\end{equation}
We note that $M_Z^2 (-\cos2\beta)/2$ is an eigenvalue of the 
matrix: ${\rm diag.}(-1,-1,1,1) M_0^2$,
and the corresponding eigenvector is $(v_{1u},v_{2u},v_{1d},v_{2d})$.
The $Z$ boson mass can be written as
\begin{equation}
\frac{M_Z^2}2 \cos^22\beta = -  \left(
\begin{array}{cccc}
 v_{1u}  & v_{2u} & v_{1d} & v_{2d}
\end{array}
\right)
M_0^2\left(
\begin{array}{c}
v_{1u}  \\ v_{2u} \\ v_{1d} \\ v_{2d}
 \end{array}
\right)\frac{1}{v^2}.
\end{equation}
The symmetry breaking condition is $\det M_0^2 < 0$.
In this case, a large $\tan\beta$ (and $\cos\zeta_u\sim1$) can be obtained
by small $m_{u12}^2$, $\hat b_{11}$ and $\hat b_{12}$. 
The symmetry breaking can arise when the determinant of the sub-matrix $(M_0^2)_{ij} (i,j = 2,3,4)$ is negative,
and it is not necessarily true that  a cancellation in $m_{u11}^2$
between $-(m^2_{H_u})_{11}$ and $\hat\mu_{11}^2 + \hat\mu_{12}^2$ needs to occur
to obtain the little hierarchy.
Surely, a cancellation is needed to make the magnitude of the determinant of $M_0^2$ small
for the little hierarchy.
The cancellation happens radiatively at $Q_0$ (by definition) and the important tuning quantity is the size of
$\ln Q_0/Q_S$.
In 4HD case, we obtain
\begin{equation}
\frac{M_Z^2}2 \cos^22\beta \simeq  \left(
\begin{array}{cccc}
 v_{1u}  & v_{2u} & v_{1d} & v_{2d}
\end{array}
\right)
\frac{d M_0^2}{d\ln Q} \left(
\begin{array}{c}
v_{1u}  \\ v_{2u} \\ v_{1d} \\ v_{2d}
 \end{array}
\right)\frac{1}{v^2} \ln \frac{Q_0}{Q_S}.
\label{MZ-RGE}
\end{equation}
In the Yukawa-basis, $d (M_0^2)_{11}/d\ln Q$ and $d (M_0^2)_{33}/d\ln Q$  are positive due to the Yukawa interaction. The size of the term $d(M_0^2)_{(11,33)}/d\ln Q$ is governed by the stop mass. $(M_0^2)_{11}$ and $(M_0^2)_{33}$ are smaller at the lower energy side as happens in the 2HD case.
However, the other component of $d (M_0^2)_{ij}/d\ln Q$ can be negative\footnote{
A careful treatment is needed since the signs of the off-diagonal elements depend on the signs of $\mu_{ij}$ and $b_{ij}$
(under the convention $0< \zeta_u,\zeta_d,\beta<\pi/2$).
}.
In fact, due to the absence of Yukawa coupling (in the Yukawa-basis by definition),
 $d (M_0^2)_{22,44}/d\ln Q$ is negative, and thus, $(M_0^2)_{22,44}$ becomes larger at the lower energy.
 This can make to keep $\det M_0^2 \approx 0$ for a wider range of $Q_S$ compared to 2HD case.
Roughly speaking, for a lighter stop mass $\sim 2-3$ TeV, if the heavier Higgsino mass is $O(10)$ TeV, one finds that the sensitivity for $\ln Q_0/Q_S$ is relaxed,
and the little hierarchy is much less fine-tuned compared to the 2HD case.

In order to illustrate the above statement,
let us rewrite Eq.(\ref{MZ-RGE}) using a bold approximation.
We neglect the terms which depend on $\cos\beta$,
and gaugino masses $M_1$ and $M_2$.
We also neglect the terms which depends on $\hat \mu_{12}$ and $\hat \mu_{21}$,
assuming that the Higgs mixing $\zeta_u$ is mainly generated by SUSY breaking term, $(\hat m_{H_u}^2){}_{12}$,
and that the dominant contribution from 2HD case is proportional to $\hat \mu_{22}^2$.
Then, we can write approximately as
\begin{equation}
\frac12 M_Z^2 \simeq 
\frac{1}{16\pi^2} \left(
 \cos^2\zeta_u (6y_t^2(m_{\tilde t_L}^2 + m_{\tilde t_R}^2) +6A_t^2)
 + \sin^2\zeta_u (- 6 g_2^2 - 2g^\prime{}^2) \hat \mu_{22}^2
\right) \ln \frac{Q_0}{Q_S}.
\end{equation}
For example,
suppose that $m_{\tilde t_L}= m_{\tilde t_R} = 2$ TeV
and $A_t = 5$ TeV.
In 2HD case (which corresponds to $\sin\zeta_u = 0$),
we obtain $\ln Q_0/Q_S \simeq 0.003$,
and it means that $m_1^2 m_2^2 - m_3^2 \approx 0$
is satisfied only in a narrow range,
and $Q_S$ needs to be fine-tuned and to be very close to $Q_0$.
The approximate relation tells us
that $\det M_0^2 \approx 0$ can be satisfied in a wide range
if the heavier Higgsino mass is chosen to be
\begin{equation}
\hat \mu_{22}^2 \sim \cot^2\zeta_u \frac{6y_t^2(m_{\tilde t_L}^2 + m_{\tilde t_R}^2) +6A_t^2}{6 g_2^2 +2g^\prime{}^2},
\end{equation}
and electroweak symmetry breaking can happen ``naturally".
One can find that
$\hat \mu_{22} \sim 20,30,50$ TeV
for $\cos\zeta_u = 0.92, 0.96, 0.98$, respectively.

We note that the wino, bino and one of the Higgsino (and one of 
the charged Higgs (as well as the CP-odd neutral Higgs)) can be light ($\sim 1$ TeV) in the 4HD scenario,
while the other one needs to be heavy to relax the sensitivity which appears in the  2HD case.

\section{Conclusion}

The doublet-triplet splitting problem is one of the major issue in the grand unified models.
The missing partner mechanism is known to provide a solution to the problem.
In principle, the number of  the pairs of Higgs doublets is a free parameter in this mechanism,
though one pair of Higgs doublets is the minimal choice and it is 
preferable for the gauge coupling unification which can have additional contributions from GUT thresholds and intermediate scales.
In this paper, 
we have investigated the possibility
that two pairs of Higgs doublets (i.e., four Higgs doublets, 4HD)
remain at the TeV scale.
In 2HD, the bottom-tau unification, which is one of the major implication of GUTs,
does not appear to be successful after including the current experimental constraints from LHC.
In fact, for a suitable parameter region of $\tan\beta$ where the RGE runnings 
allows us to generate bottom-tau unification,
large threshold corrections from SUSY breaking are generated which ruin this unification.
In 4HD, on the other hand,
we find that the threshold corrections can be small 
even if the tree-level Yukawa coupling associated with   bottom and tau are
unified by RGE.
This happens due to  the freedom of the Higgs mixing terms at the TeV scale. It is possible to choose two pairs of Higgs doublet to be light and 
a linear combination of these two  pairs acquire the vacuum expectation values by the minimization of the Higgs potential.
The top-bottom-tau and bottom-tau Yukawa unifications are also implied in the context of SO(10) and SU(5) models respectively   in this scenario.
We also discuss the merits of 4HD compared to the 2HD choice
for the little hierarchy between the SUSY breaking masses and the $Z$ boson mass.
The additional Higgs pair at $O(10)$ TeV appears to relax the sensitivity of the radiative electroweak symmetry breaking. 

\section*{Acknowledgments}

%\noindent

The work of B.D. is supported by DOE Grant de-sc0010813.
The work of Y.M. is supported by grant MOST
106-2112-M-002-015-MY3
of R.O.C. Taiwan,
and 
Scientific Grants 
by the Ministry of Education, Culture, Sports, Science and Technology of Japan 
(Nos. 16H00871, 16H02189, 17K05415 and 18H04590).

\end{document}